\documentclass{aa}
\usepackage{amssymb}
\usepackage{latexsym}
\usepackage{graphicx}
\usepackage{longtable,lscape,graphicx}
\usepackage{amsmath}

\begin{document}

\title{On the effect of the inhomogeneous subsurface flows\\
on the high degree solar $p$-modes}
\author{B.M.~Shergelashvili\thanks{On leave from Center
for Plasma Astrophysics, Abastumani Astrophysical Observatory, 2a
Kazbegi Ave., Tbilisi 380060, Georgia} \and S.~Poedts}

\institute{Centre for Plasma Astrophysics, Katholieke Universiteit
Leuven, Celestijnenlaan 200B, 3001 Leuven, Belgium}

\date{Received; accepted}
\abstract{The observed power spectrum of high-degree solar
$p$-modes ($\ell>200$) shows discrepancies with the power spectrum
predicted by the stochastic excitement and damping theory. In an
attempt to explain these discrepancies, the present paper is
concerned with the influence of the observed subsurface flows on
the trapped acoustic modes ($p$-modes). The effect of these
inhomogeneous background flows is investigated by means of a
non-modal analysis and a multi-layer model. It is shown that the
rotational and meridional components of the velocity field change
the wavelengths of the oscillation modes which, in turn, results
in modifications of the corresponding modal frequencies. The
magnitudes of the frequency residuals depend on the spatial scales
of the modes and on the gradients of the different components of
the flow velocity. Together with other mechanisms (e.g.\ the
scattering of modes by the large scale convection (Goldreich \&
Murray \cite{gom})), the non-modal effect of the variation of the
frequencies in time may contribute: 1)~to the observed widening of
the corresponding peaks in the observed power spectrum with
increasing angular degree; 2)~to the partial dissipation of
spectral power, and, as a result, 3)~to the discrepancies between
the predicted and the observed power spectrum of solar $p$-modes.

}

\titlerunning{The effect of subsurface flows on $p$-modes}
\authorrunning{B.M.~Shergelashvili and S.~Poedts}

\maketitle

\section{Introduction}

One of the main goals of helioseismology is to understand the
excitation and damping mechanisms that yield the observed power
spectrum of the solar oscillation modes. The theory of the random
generation of modes by convective turbulence has been developed by
Goldreich et al.\ (\cite{gomur}) and Goldreich \& Kumar
(\cite{gokum}) in an attempt to quantitatively understand the
structure of mode sources and the generation rate of the
oscillation modes. Observations show (see e.g. Libbrecht
(\cite{lib}); Woodard et al.\ (\cite{wo})) that the energy of the
$p$-modes with frequencies below the photospheric cut-off grows
with the angular degree $\ell$ for intermediate values of $\ell$,
i.e.\ $100<\ell<200$. This observational fact is in good agreement
with the predictions from the theory of turbulent excitement and
damping of $p$-modes mentioned above. However, for $\ell>200$
there are {\em substantial discrepancies between the theory and
the observations}. As a matter of fact, recent observations by
Woodard et al.~(\cite{wo}) show that for $\ell>200$ the wave
energy distribution function {\em decreases rapidly with
increasing angular degree} $\ell$. The latter authors remark that
the discovered decline in mode energy at high $\ell$ values may
indicate the existence of some `unmodeled mechanism of damping'.
In the present paper, the viability of such a mechanism is
studied, viz.\ mode damping due to {\em a sheared background
flow}. As a matter of fact, there exist numerous
helioseismological measurements of solar subsurface flows (see \
Gonz\'alez Hern\'andez et al.\ (\cite{gprdia}), Gonz\'alez
Hern\'andez \& Patr\'on J. (\cite{gons})). These observations show
that the velocity field of these flows has both rotational and
meridional components and that the gradients of these flow
velocity components depend on the spatial coordinates. We will
present a multi-layer model for these flows that allows to
quantify their effect on the solar $p$-modes.

Several other candidate mechanisms for non-turbulent wave damping
have been suggested, e.g.\ the nonlinear coupling between trapped
and propagating modes (Kumar \& Goldreich (\cite{ku})) or mode
scattering by convective motions (Goldreich \& Murray \cite{gom}).
Moreover, the direct influence of the inhomogeneous temperature,
the magnetic field, and/or the velocity fields on the solar
$p$-modes and $f$-modes have been studied intensively (Murawski \&
Roberts (\cite{mu1}, \cite{mu2}); Murawski \& Goossens
(\cite{mugo}); Vanlommel \& \v{C}ade\v{z} (\cite{va1}), Vanlommel
\& Goossens (\cite{va2})), while `resonant absorption' (or
`continuum damping') has been introduced as a candidate damping
mechanism for $p$-modes by Tirry et al.~(\cite{ti}). Clearly, the
physical characteristics of the solar medium affect and
systematically distort the main frequencies and the travel times
of the modes. This insight has led to the development of the
time-distance technique in helioseismology (Duval et al.\
(\cite{du})) which became a powerful tool for the observation of
3D structures with inhomogeneous magnetic and temperature
profiles.

As pointed out by Goldreich et al.\ (\cite{gomur}), when the
widths of the frequency peaks in the power spectrum are greater
than the inverse of the observation time, these peaks contain
information about the damping rates of the modes. These authors
also indicated that the excess width of the peaks (with respect to
the inverse observation time) can result from a variation of
either the frequency or the amplitude of the modes and that it is
reasonable to assume that both these types of velocity modulation
yield comparable contributions in the line width. Observations
show that the line widths of the oscillation modes increase with
increasing frequency and angular degree. Goldreich \& Murray
(\cite{gom}) suggested the scattering of modes by turbulent
velocity fluctuations as a possible mechanism supporting this
observational fact.

It is well known that the main properties of the solar oscillation
modes are well described by the standard normal mode analysis. For
a static equilibrium and for appropriate boundary conditions, the
linear MHD (magnetohydrodynamics) operator is self-adjoint.
However, in neutral fluids and in plasmas with {\em nonuniform
background flows} there are often discrepancies between the
analytical results of the standard modal analysis and experimental
data. These discrepancies are usually ascribed to the mathematical
incompleteness of the normal mode spectrum due to the
non-self-adjoint nature of the governing equations (Trefethen et
al. (\cite{tre}), Criminale \& Drazin (\cite{crdr2}),
(\cite{crdr1})). One of the alternative and complementary methods
to solve the initial value problem in sheared plasma flows is the
{\em non-modal analysis}. In this formalism a transformation of
the variables is performed and, as a result, the spatially
inhomogeneous terms in the governing equations are replaced by
time-dependent terms. Therefore, in general, this method enables
to study the non-exponential temporal evolution of linear
perturbations of sheared flows. The non-modal analysis was
introduced originally by Lord Kelvin (\cite{kel}) and the
theoretical basis for studying the wave dynamics in hydrodynamical
and plasma flows has been developed by Goldreich \& Linden-bell
(\cite{gol}), Craik \& Criminale (\cite{cr}); Chagelishvili et
at.~(\cite{cha1}); Rogava et al.~(\cite{ro2,ro1}). Many authors
showed that the shear in the flow velocity field can have
substantial effects on the wave dynamics, e.g.\ causing
time-dependent frequencies and wave numbers (Butler \& Farrell
(\cite{bufar}); Chagelishvili et al. (\cite{charseg})) and
coupling of different wave modes (Chagelishvili et al.
(\cite{chartsi}); Rogava \& Mahajan (\cite{rogmah})). The
non-modal approach has also been applied to the study of several
aspects of the excitement of gravitational and acoustic waves on
the Sun (see Chagelishvili et al.\ (\cite{cha2}); Pataraya \&
Pataraya (\cite{pat}).

The significant discrepancies between the observed and predicted
power spectra of trapped modes with high angular degree $\ell>200$
may be due to sheared flow effects. As a matter of fact, these
modes mostly propagate in the thin cavity below the photosphere
($r>0.95R_{\sun}$) in which plasma flows with strongly pronounced
velocity gradients have been observed. Hence, the aim of our study
was to investigate the possible role of the non-exponential
temporal evolution of the modes (due to the nonuniform flows) in
the formation of the observed power spectra. We applied a
non-modal analysis to study the effect of the observed
sub-photospheric shear flows on the frequencies and the
propagation characteristics of the solar $p$-modes. In order to
distinguish the effect of velocity shear from all other effects,
we adopt a very simple model and consider the modes locally, in
plane slab geometry. Our model allows to separate the effects
related to the spatial inhomogeneity of the background flow from
the inhomogeneity coming from the gravitational stratification and
the temperature gradient and enables us to focuss on the evolution
of the wave characteristics associated with the flow inhomogeneity
itself.

In the next section, we present the model we applied to
investigate the interaction between the inhomogeneous flows
beneath the photosphere and trapped acoustic modes. In the third
section, we describe and analyze our theoretical results. In the
fourth section, we discuss the obtained numerical results and we
present and discuss some phenomenological points of view in the
framework of the obtained results. Then, we discuss the possible
observational consequences of the shear flow effects on the $p$-modes.
In the final section, we formulate our conclusions.

\section{Physical model and equations}
The aim is to study the temporal evolution of small perturbations
in a stratified medium with a non-uniform two-dimensional
`background' flow. The problem is considered in plane-parallel
geometry and is described in Cartesian coordinates. The $x$-axis
denotes the `toroidal' direction and its positive direction is
chosen as the direction in which the Sun rotates. The $y$-axis
coincides with the `meridional' direction and is directed towards
the north pole. The $z-$axis covers the `radial' direction and is
directed outwardly, to the solar atmosphere. We study the
properties of modes with small amplitudes which are considered as
linear perturbations of the equilibrium state.

\subsection{The equilibrium model}
\begin{figure}
\centering \includegraphics[width=7.5cm]{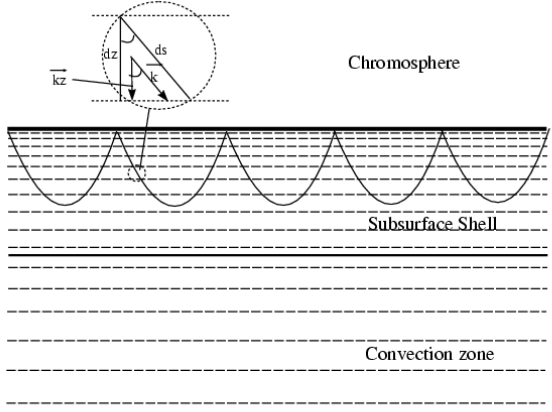}
\caption{ Schematic view of the modelled subsurface spherical
shell (in plain slab geometry). The dashed horizontal lines
schematically represent the boundaries of the thin layers
considered in our multi-layer model. Within each of these thin
layers the sound speed is assumed to be approximately constant and
they correspond to the piecewise constant temperature profiles
shown in Figure \ref{proftemp1} or \ref{proftemp2}.} \label{f1}
\end{figure}
In this subsection, the main parameters characterizing the
equilibrium are introduced, in particular the parameters
characterizing the equilibrium flow velocity. These parameters are
determined in such a way that they yield a good approximation of
the above-mentioned measurements of the solar subsurface flow
profiles. In general, the subsurface flows can be very
complicated. In the non-modal analysis, the nine components
describing the shear rates of the background flow velocity $\vec
V_{0}$ components in the three spatial directions are collected in
the so-called {\em `Shear Matrix'} $S$ (for details e.g.\ see
Mahajan \& Rogava \ (\cite{mahro})). In its most general form this
matrix reads:
\begin {equation}\label{shmg}
S\equiv \left( \begin{array}{ccc}
V_{0x,x} & V_{0x,y} & V_{0x,z} \\
V_{0y,x} & V_{0y,y} & V_{0y,z} \\
V_{0z,x} & V_{0z,y} & V_{0z,z}
\end{array}\right ),
\end{equation}
where the index after the comma denotes the derivative with
respect to the corresponding spatial coordinate. We want to study
the joint influence of the rotational and the meridional
subsurface flows on the solar $p$-modes. Hence, the equilibrium
velocity field includes both rotational ($x-$) and meridional
($y-$) components. In our general model we consider the
$x$-component of the equilibrium flow velocity as a function of
both the $y$- and $z$-coordinates, $V_{0x}=V_{0x}(y,z)$. Further,
we also take into consideration the meridional component of the
flow velocity and its dependence on the radial coordinate, i.e.\
$V_{0y}=V_{0y}(z)$. Consequently, in the equilibrium state we have
an inhomogeneous velocity field $\vec{V}_0 =
(V_{0x}(y,z),V_{0y}(z),0)$.

In section~4 we will model the subsurface flows with a linear
approximation of the velocity profiles measured by Gonz\'alez
Hern\'andez \& Patr\'on (\cite{gons}). Hence, in the local frame
of reference the components of the flow velocity read as:
\begin {equation}\label{vx0case3}
V_{0x}=V_{01}+ay+bz,
\end{equation}
and
\begin {equation}\label{vy0case3}
V_{0y}=V_{02}+cz,
\end{equation}
where, $V_{01}$ and $V_{02}$ represent the constant parts of the
rotational and meridional components of the flow velocity and the
parameters $a=V_{0x,y}$, $b=V_{0x,z}$ and $c=V_{0y,z}$ are the
local spatial derivatives of the corresponding velocity component.
Note, that the considered background flow is incompressible
$\nabla\cdot\vec{V}_0=0$.

In order to consider the propagation of small perturbations in the
gravitationally stratified medium we adopt a local analysis,
similar to the formalism used by Whittaker (\cite{whit}) and
Ulrich (\cite{ul}). It should be noted, however, that we here
focuss on the effects of the inhomogeneous background flow and
thus we excluded the effects related to the magnetic field,
radiative flux and convective motions. The latter issue has
already been addressed by several authors (e.g. see Swisdak and
Zweibel (\cite{swzw})). In addition, the influence of the random
velocity fields on $f$-modes has been studied by Murawski \&
Roberts (\cite{mu1}, \cite{mu2}) and by Murawski \& Goossens
(\cite{mugo}) also including the effect of the chromospheric
magnetic field. Within each of the thin shells in our model
(schematically shown by the horizontal dashed lines in Figure~1
the temperature (i.e.\ sound speed) is assumed to be constant.
Hence, the temperature profile of the standard solar model (see
e.g.\ \cite{solmod}) is approximated by a step-function. In other
words, locally, in each thin layer we have an equilibrium density
and pressure profile of the form $\rho _0(z),p_0(z)\propto
\exp{\left (-z/H \right )}$,
where $H$ is the local scale height, so that in each layer the
temperature is constant.

\subsection{The temporal evolution of the $p$-modes}
Let us now study the propagation of small perturbations in the
considered plasma shell and focuss on the effects related to the
inhomogeneity of the background flow(s). We consider the waves as
{\em linear} perturbations of the stationary equilibrium described
in the previous section. The following set of linearized equations
then governs the dynamics of the small disturbances in a
gravitationally stratified isentropic gas:
\begin{subequations}\label{eqsl}
\begin{equation}\label{eqcl}
\frac{\partial\rho_{1}}{\partial
t}+(\vec{V}_{0}\mathbf{\cdot}\vec{\nabla
})\rho_{1}+(\vec{v}_1\mathbf{\cdot}\vec{\nabla})\rho_{0}(z)+\rho_{0}%
(z)(\vec{\nabla}\mathbf{\cdot}\vec{v}_1)=0,
\end{equation}
\begin{displaymath}
\rho_{0}(z)\frac{\partial\vec{v}_1}{\partial t}+\rho_{0}(z)\left(  \vec{V}%
_{0}\cdot\vec{\nabla}\right)  \vec{v}_1+\rho_{0}(z)\left(
\vec{v}_1\cdot \vec{\nabla}\right) \vec{V}_{0}=\quad
\end{displaymath}
\begin{equation}\label{eqmol}
 \qquad\qquad\qquad\qquad\qquad -\vec{\nabla}p_{1}+\rho_{1}\vec{g},
\end{equation}
\begin{displaymath}
\frac{\partial p_{1}}{\partial t}+\left(
\vec{V}_{0}\cdot\vec{\nabla}\right)
p_{1}+(\vec{v}_1\mathbf{\cdot}\vec{\nabla})p_{0}(z)=\qquad\qquad\qquad\qquad
\end{displaymath}
\begin{equation}\label{eqel}
\qquad\qquad\frac{\gamma p_{0}}{\rho _{0}}\left(
\frac{\partial\rho_{1}}{\partial t}+\left(  \vec{V}_{0}\cdot
\vec{\nabla}\right)
\rho_{1}+(\vec{v}_1\mathbf{\cdot}\vec{\nabla})\rho _{0}(z)\right),
\end{equation}
\end{subequations}
where $\vec{g}$ denotes the gravitational acceleration and
$\gamma$ is the ratio of specific heats, symbols with index 0
denote the equilibrium quantities and those with index 1 the
perturbations.

Clearly, in the case of a static medium, i.e.\ without the
background flows, we get a simplified set of equations without the
inhomogeneous terms coming from the convective derivatives. In
this case, the normal-mode solution of this set of equations takes
the form:
\begin{equation}\label{solh}
\psi \sim \exp \left [ i\left (\omega t + k_xx+k_yy +\left(
k_{z}-\frac{i}{2H}\right)z \right ) \right ],
\end{equation}
where, $\psi $ represents the quantities $\rho _{1}/\rho _{0}$,
$p_{1}/\rho_{0}$ and the components of the velocity perturbation.
However, we want to know what happens to these solutions when the
non-uniform equilibrium flows are taken into account. Notice that
Eqs.~(\ref{eqsl}) contain convective derivatives, which also take
into account the spatial inhomogeneity of the initial
(equilibrium) velocity field. These spatially inhomogeneous terms
complicate the analysis considerably. For small magnitudes of the
background flow and of its gradients the problem has been studied
in the standard framework of the normal modal analysis and the
influence of the flow on the eigenfrequencies has been treated as
a small perturbation of the basic frequencies. This perturbation
technique is applied in Backus and Gilbert (\cite{bagil}) and
Ulrich et al.~(\cite{ulr}). The methods of inversion of the solar
internal rotation rate, from the frequency splitting observations
of the solar non-radial oscillations, are based on a similar
theoretical formalism. These methods allow to determine the
internal rotational velocity profile of the Sun with a high
accuracy.

However, due to the inhomogeneous background flow the governing
equations have a non-self-adjoint character. Hence, we apply the
non-modal technique, which enables us to find an additional class
of `non-normal' solutions of the above set of
equations~(\ref{eqsl}). These solutions describe the evolution of
the perturbations in time which, in general, can be
non-exponential.

Following the non-modal technique we represent the perturbation
quantities $\rho_1/\rho_0$, $p_1/\rho_0$, and $\vec{v}_1$ in the
following form:
\begin{equation}\label{psim}
\psi(x,y,z;t)=\hat{\psi} (k_x,k_y,k_z;t)\exp{\left ( i \left
(\varphi_1-\varphi_2-\frac{i z}{2H}\right )\right )},
\end {equation}
where,
\begin {equation}\label{phi1}
\varphi_1 (x,y,z;t)=k_x(t)x+k_y(t)y+k_z(t)z,
\end{equation}
and
\begin {equation}\label{phi2}
\varphi_2 (k_x(t),k_y(t);t)=
V_{01}\int{k_x(t)dt}+V_{02}\int{k_y(t)dt}.
\end{equation}
In these expressions $k_x(t)$, $k_y(t)$ and $k_z(t)$ are the (in
general) time dependent components of the wavevector satisfying
the following set of ODEs:
\begin {equation}\label{eqk}
\frac{d\vec{k}}{dt}+S^{\rm T}\cdot \vec{k}=0,
\end{equation}
where $S^{\rm T}$ is the transposed shear matrix. In this general
formalism, the substitution of expressions
(\ref{psim})-(\ref{phi2}) in the governing equations results in a
set of {\em ordinary} differential equations (ODEs) in time in
which the spatial inhomogeneity is replaced by a temporal one (for
details see e.g.\ Rogava et al.~(\cite{ro1})). These ODEs govern
the temporal evolution of the spatial Fourier harmonics with
time-dependent components of the wave vector. In the case of the
two dimensional non-uniform flow considered here, one gets:
\begin{subequations}\label{eqksol}
\begin{equation}\label{eqkxsol}
\frac {dk_x}{dt}=0,
\end{equation}
\begin{equation}\label{eqkysol}
\frac {dk_y}{dt}+ak_x=0,
\end{equation}
\begin{equation}\label{eqkzsol}
\frac {dk_z}{dt}+bk_x + ck_y=0.
\end{equation}
\end{subequations}
Consequently, for the considered flow velocity profiles the
$x$-component of the wave vector remains constant in time,
$k_x=k_{x0}$, but
\begin{equation}\label{skysf}
k_y(t)=k_{y0}-ak_xt.
\end{equation}
Integrating Eq.~(\ref{eqkzsol}), we find for the temporal
evolution of the radial ($z$-) component of wavevector:
\begin{equation}\label{skzsf}
k_z(t) = k_{z0} - (bk_{x0}+ck_{y0})\,t+\frac{ack_{x0}}{2}\,t^2.
\end{equation}

It should also be noticed that (Rogava (\cite{rogava2002})) there
exists a combination of the solutions of the
Eqs.~(\ref{eqkxsol})-(\ref{eqkzsol}) that is preserved in time,
viz.\
\begin {equation}\label{delta}
\Delta \equiv ck_y ^2+2(bk_yk_x-ak_zk_x)=\hbox{const}.
\end{equation}
Using the representation (\ref{psim}) in Eqs.~(\ref{eqsl}) and
rewriting the equations for the components of the velocity
perturbations, normalized density perturbation $D=\rho_1/\rho_0$
and the enthalpy perturbation $Q=p_1/\rho_0$)
we get the following set of ODEs for the perturbed quantities:
\begin{subequations}\label{eqwnon}
\begin{equation}\label{eqw1non}
\frac{d\hat{D}}{dt}+i\left( k_{x}\hat{v}_{x}+k_{y}(t)\hat
{u}_{y}+\tilde{k}_{z}(t)\hat{v}_{z}\right) =0,
\end{equation}
\begin{equation}\label{eqw2non}
\frac{d\hat{v}_{x}}{dt}+a\hat{v}_{y}+b\hat{v}_{z}=-ik_{x}\hat{Q},
\end{equation}
\begin{equation}\label{eqw3non}
\frac{d\hat{v}_{y}}{dt}+c\hat{v}_{z}=-ik_{y}(t)\hat{Q},
\end{equation}
\begin{equation}\label{eqw4non}
\frac{d\hat{v}_{z}}{dt}=-i\tilde{k}_{z}(t)\hat{Q}+g\hat{D},
\end{equation}
\begin{equation}\label{eqw5non}
\frac{d\hat{Q}}{dt}=C_{s}^{2}\left( \frac{d\hat{D}}{dt}
+\frac{(\gamma-1)\hat{v}_{z}}{\gamma H}\right),
\end{equation}
\end{subequations}
where $\tilde{k}_{z}(t)=k_{z}(t)+i/2H$. Hence, the presence of
shear flows causes a transformation of the wave solutions in time. The
rate of this transformation strongly depends on the initial
orientation of the wave vector and on the values of the shear
parameters included in the shear matrix.

Our first aim is to {\em qualitatively} understand the temporal
behavior of the $p$-modes driven by the joint effect of the
inhomogeneous rotation and the meridional flow appearing in the
thin shell just below the photosphere. Therefore, in the next
section we will introduce some simplifying assumptions. This will
enable us to derive an approximate analytical `dispersion
relation' which governs the  temporal evolution of the `effective
frequencies' of the modes. We will also demonstrate the shear flow
effect in a simple model with a linear velocity profile, viz.\ for
a mode with $l = 951$ which is confined within a very thin cavity
above the fractional radius 0.997. In section~4 we then try to
{\em quantify} this effect in first order with a multi-layer
model. The measured velocity profiles are approximated linearly in
each layer and the cumulative effect of the layers is calculated
taking into account the time spent by the modes in each layer.

\section{Approximative analysis}
\begin{figure}
\centering
\includegraphics[width=8cm]{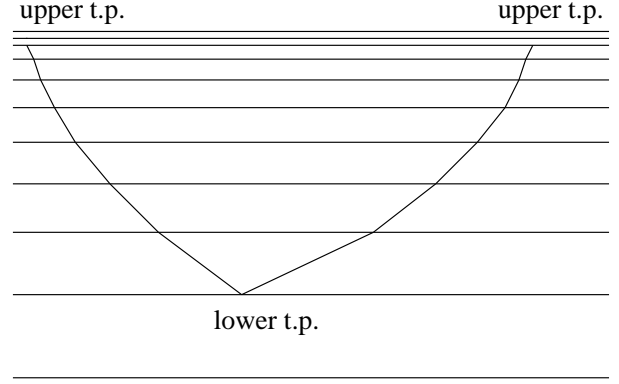}\\
  \caption{Schematic view of the piecewise straight ray path.}\label{ray}
\end{figure}

In general, the application of the non-modal approach means that,
in the governing equations, the spatial inhomogeneity (related to
the terms containing the nonuniform background flows) is replaced
by a temporal one. After this transformation of the perturbation
variables the temporal behavior of the spatial Fourier harmonics
(containing the time dependent wavenumbers) can be studied. In
most cases, the solution of the resulting ODEs (in time) requires
the application of numerical techniques. In this particular case
of the trapped acoustic-gravity modes, the initial value problem
has a specific feature. On one hand, the time dependence of the
wavevector results in a variation of the phase speed and the
direction of the wave propagation, in time. On the other hand, the
temperature gradient refracts the propagating mode. Formally
speaking, the local approach outlined in the previous section is
the `non-modal equivalent' of the so called ray approximation,
commonly used for the local description of $p$-modes (Whitaker
(\cite{whit}); Ulrich (\cite{ul})). But, in this case we do not
perform a Fourier transform with respect to time. Instead, one has
to study the propagation of the modes by solving the governing
equations (\ref{eqwnon}) in each layer separately and then
matching the solutions at the interfaces between the different
layers. However, at this stage we aim to qualitatively estimate
the characteristic changes in the modal properties in order to
understand the principle of the mechanism of these changes induced
by the inhomogeneity of the background flows. Therefore, we adopt
an approximative analysis to study the propagation properties of
the modes and to construct the shape of their ray paths. The part
of the ray path confined within a given layer, with constant
temperature, is locally linear (a schematic view of this kind of
piecewise straight ray path is shown in Figure~\ref{ray}) and
within this layer the equations governing the motion of a point of
the wave front can be written as follows:
\begin{subequations}\label{eqph}
\begin{equation}\label{eqphx}
    \left ( \frac {\delta x}{\delta t} \right ) _i=v_{ph}(z_i,t)\frac{k_{x_i}}{|k_i|}=
\frac{\omega (t)k_{x_i}(z_i,t)}{|k_i(z_i,t)|^2},
\end{equation}
\begin{equation}\label{eqphy}
   \left (  \frac {\delta y}{\delta t} \right ) _i=v_{ph}(z_i,t)\frac{k_{y_i}}{|k_i|}=
    \frac{\omega (t)k_{y_i}(z_i,t)}{|k_i(z_i,t)|^2},
\end{equation}
\begin{equation}\label{eqphz}
   \left (  \frac {\delta z}{\delta t} \right ) _i=v_{ph}(z_i,t)\frac{k_{z_i}}{|k_i|}=
    \frac{\omega (t)k_{z_i}(z_i,t)}{|k_i (z_i,t)|^2},
\end{equation}
\end{subequations}
where, index $i$ denotes a given layer between horizontal planes
located at $z_{1i}=z_i$ and $z_{2i}=z_i+(\delta z) _i$; $k_i$,
$k_{x_i}$, $k_{y_i}$ and $k_{z_i}$ are the module and components
of wavevector, respectively. In these equations we introduce the time
dependent `effective' frequency $\omega (t)$. We will turn to the
issue of validity of such a description in the following subsection.
It should be noticed that, because of the temperature
gradient the local values of the sound speed in each neighboring
layer are different. Therefore, the components of the wavevector
should be adjusted accordingly, in order to keep the `effective'
frequency unchanged through the interface between the layers. One
can easily see that in the case without a shear flow the
`effective' frequencies should be replaced by the usual constant modal
frequencies. In this latter case the problem always can be reduced
to two dimensions (as it is usually done under the normal mode
formalism). In our case, however, the trajectories (ray paths) of the
modes are, in fact, 3D curves. From this point our problem is
reduced to the determination of the `effective' frequency and how
it relates to the time dependent components of the wavevector.
\subsection{The `effective' frequency}
Generally speaking, the effect of an inhomogeneous flow on waves
manifests itself in a temporal variation of the modal properties.
Using different analytical methods (among which the non-modal
approach) it has been shown that strong gradients in the
background flow velocity can cause very rapid changes of the modal
frequency and the amplitude, even in time intervals as short as an
oscillation period. As a result, the power spectrum of the waves
is affected in such a way that it may become impossible to
distinguish the individual peaks in the spectrum.

In the Sun, however, such strong gradients of the flow velocity
are not observed. Therefore, in this first attempt to model this
`small' effect {\em qualitatively}, we assume that the sheared
flows below the surface of the Sun just slightly transform the
oscillation modes and that this transformation is represented by a
small deviation of the mode properties found by the standard modal
analysis. As a result of the flow inhomogeneity there appear terms
of the form $S_{ij}u_{i}$ (where a summation over repeated indices
is meant) in the equations (\ref{eqw2non})-(\ref{eqw3non}). Here
the indices $i,j=1,2,3$ represent the $x-$, $y-$, and
$z-$direction, respectively, and $S$ is the shear matrix. From
Table~1 it is clear that the gradients of the flow velocity
components are two to four orders smaller than the characteristic
angular frequencies of the oscillation modes. Hence, we expect
that the contribution from the $S_{ij}u_{i}$ terms in the solution
of the governing equations is very small. Therefore, at this stage
we neglect these terms, at least in equations
(\ref{eqw2non})-(\ref{eqw3non}).

\begin{figure*}
  \includegraphics[scale=1]{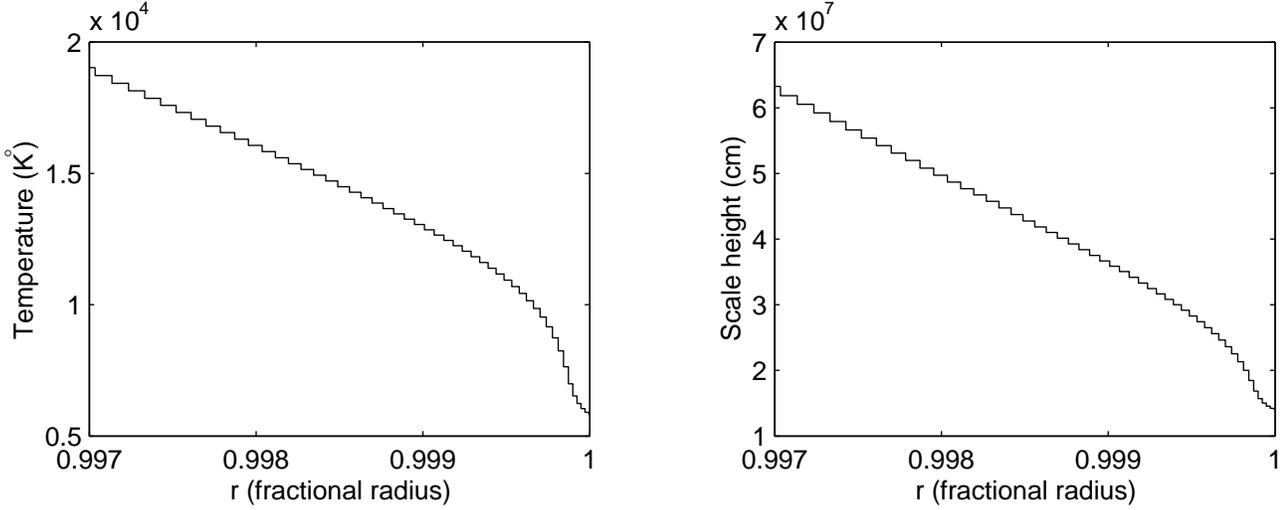}\\
  \caption{Piecewise constant radial profile of the temperature (panel~A) and
  scale height (panel~B) corresponding to the case of the simple velocity
  profile addressed in subsection~3.2.}\label{proftemp1}
\end{figure*}

\begin{figure*}
  \includegraphics[width=17cm]{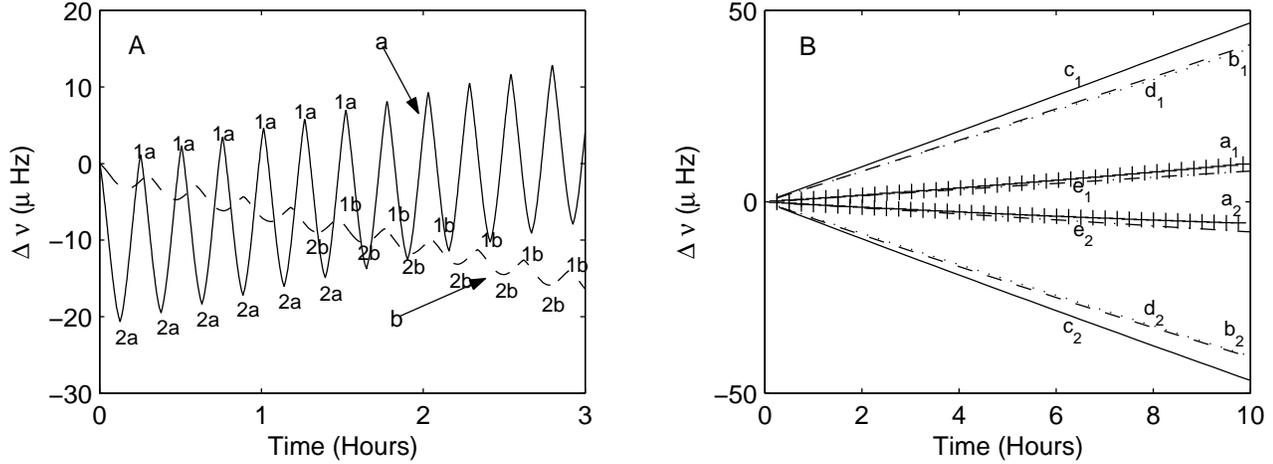}
  \caption{The variation of the modal frequency in time. The curves
correspond to the mode $\ell=951$, $n=2$ and with the basic modal frequency
 $\nu _0 =4.5243\,$mHz. The calculations have been done for a simple,
linear profile of the flow velocity with the shear parameters
equal to $a=-1\cdot 10^{-6}\,$s$^{-1}$, $b=2.5\cdot
10^{-5}\,$s$^{-1}$ and $c=2.0\cdot 10^{-5}\,$s$^{-1}$. Panel~A:
curve $a$ corresponds to  $k_{y0}>0$ (angle $\beta = 45^{\circ}$)
and curve $b$ corresponds to $k_{y0}<0$ (angle $\beta =
-45^{\circ}$). The points denoted by $1a$ and $1b$ correspond to
the frequency of the respective mode close to its upper turning
points (i.e.\ the frequencies observable on the surface!) while
the inner turning points are denoted by $2a$ and $2b$. Panel~B:
curves of the frequency variation corresponding to the upper
turning points for different directions of the horizontal
wavevector $k_h$ (i.e.\ for different values of the angle
$\beta$). The curves are denoted respectively as: $a_1$ and $a_2$
-- $\beta = \pm 5^{\circ}$ (crossed lines), $b_1$ and $b_2$ --
$\beta = \pm 30^{\circ}$ (dashed lines), $c_1$ and $c_2$ -- $\beta
= \pm 45^{\circ}$ (solid lines), $d_1$ and $d_2$ -- $\beta = \pm
60^{\circ}$ (dotted lines), $e_1$ and $e_2$ -- $\beta = \pm
85^{\circ}$ (dash-dotted lines).}\label{figsimple}
\end{figure*}

The above approximation results in a mathematical simplification
of the problem as it makes the frequency purely real, i.e.\
oscillatory. In other words, it is equivalent to the assumption
that the amplitude of the mode varies much slower in time than the
rapid oscillation itself so that the amplitude remains
approximately constant. This approximation thus results in
perturbations of the form $\hat \psi \approx \hat
\psi(t_0)exp(i\varphi(t))$. The set of Eqs.~(\ref{eqwnon}) then
leads the dispersion equation for ${\partial\varphi(t)}/{\partial
t}$, which has a similar form as the one given by Whitaker
(\cite{whit}) for waves propagating in a static medium:
\begin{equation}\label{deq}
\left(  \frac{\partial\varphi(t)}{\partial t}\right) ^{4}-\left(
\frac{\partial\varphi(t)}{\partial t}\right) ^{2}\left(
C_{s}^{2}\left( k_{h}^{2}+k_{z}^{2}\right) +\omega_{a}^{2}\right)
+\omega_{b}^{2}C_{s}^{2} k_{h}^{2}  =0,
\end{equation}
where, $k_{h}^2=k_{x}^2+k_{y}^2$, $\omega_{a}=\gamma g/2C_{s}$ is
the local acoustic cutoff frequency, and
$\omega_{b}=\sqrt{g(\gamma-1)/\gamma H}$ is the local
Brunt-V\"{a}is\"{a}l\"{a} frequency. The difference between
Eq.~(\ref{deq}) and the dispersion equation for `homogeneous' (in
the sense of flows) waves is that now the coefficients in the
equation are time dependent. Thus by using Eq.~(\ref{psim}) we can
write the approximate `effective frequency' as the time derivative
of the time dependent phase:
\begin{displaymath}
    \omega = V_{01}k_{x} +V_{02}k_{y}-y\frac{\partial k_{y}}{\partial t}
-z\frac{\partial k_{z}}{\partial t} - \frac{\partial
\varphi(t)}{\partial t}=
\end{displaymath}
\begin{equation}\label{dispr}
=(\vec k\cdot \vec V_{0})- \frac{\partial \varphi(t)}{\partial t},
\end{equation}
where,
\begin{equation}\label{sig}
\frac{\partial \varphi(t)}{\partial t}=-\sqrt{\frac{1}{2} W
+\frac{1}{2}\sqrt{W^{2}-4\omega_{b}^{2} C_{s}^{2}k_{h}^{2}(t)}},
\end{equation}
\begin{equation}\label{W}
    W=C_{s}^{2}\left(  k_{h}^{2}(t)+k_{z}(t)^{2}\right)
    +\omega_{a}^{2}.
\end{equation}

In Eq.~(\ref{sig}) we chose the $'-'$ sign before the square root
without loss of generality. It can easily be shown that the first
four terms in the RHS of Eq.~(\ref{dispr}) in fact represent the
expression $(\vec k\cdot \vec V_{0})$. It is well-known that this
scalar product of the wavevector and the background flow velocity
represents the effect of the Doppler shift of the frequencies due
to the fact that the waves are advected by the flow. Hence, this
term is analogous to the one given for example by Ulrich et al.\
(\cite{ulr}). However, here the components of the wavevector (and
thus the characteristic frequency) are time dependent. This time
dependency represents the deviation from a purely exponential
($\sim \exp (i\omega t)$) temporal evolution of the small
disturbances.
\subsection{Effect of a simple (linear) velocity profile}
\begin{figure}
  \includegraphics[scale=1]{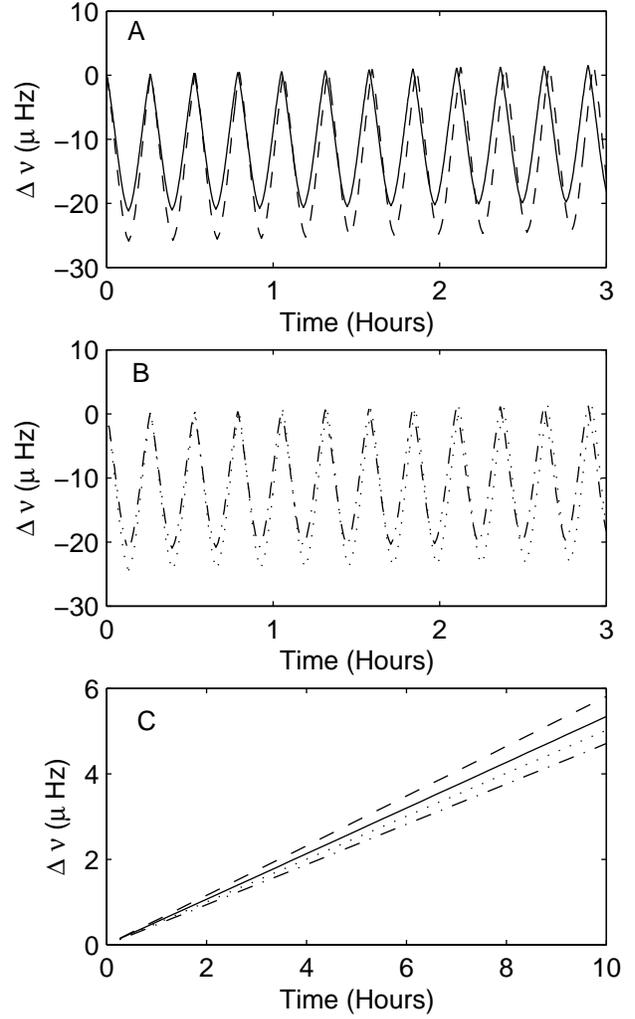}
  \caption{Results of the calculations for the same mode as
in Fig.~\ref{figsimple} for a smaller value of the shear parameter
$a=-0.1\cdot 10^{-6}$ s$^{-1}$. The value of the shear parameter
$b$ has been taken the same as in Fig.~\ref{figsimple}
($b=2.5\cdot 10^{-5}\,$s$^{-1}$). In panel~A we show the curve of
the frequency variation in time, for the angle $\beta=45^{\circ}$,
within the first three evolution hours. The solid line corresponds
to $c=2.0 \cdot 10^{-5}\,$s$^{-1}$ and the dashed line to $c=3.0
\cdot 10^{-5}\,$s$^{-1}$. Similar curves are plotted in panel~B
for the angle $\beta =30^{\circ}$, where the dashed-dotted line
corresponds to the smaller value of the parameter $c$ and the
dotted line to the larger one. Panel~C shows the resulting modal
frequency evolution over a time interval of 10 hours. The
different line styles correspond to the line styles in the
previous panels.}\label{figsimsmall}
\end{figure}

In order to numerically calculate the changes of the modal
frequencies purely due to the non-modal effects we developed a
numerical code making use of a spatial discretisation on the basis
of the Standard solar model ({\tt
www.ap.stmarys.ca/$\sim$guenther/solar/ssm455.sink}) and the
approximate `dispersion relation' (Eq.~(\ref{dispr})). In our
calculations we used also the observed $p$-mode frequency data
(published online by Libbrecht et al.\ and Bachmann et al., see
the web page {\tt http://www.gong.noao.edu/teams/data/
jwl\_freqs.html}). As an example we examined the mode with $\ell =
951$. In Figure~\ref{proftemp1} we show the piecewise constant
radial profiles of the temperature (panel~A) and scale height
(panel~B). The shown range of the fractional radius includes the
very thin cavity in which the $\ell = 951$ mode is confined.

Notice that these illustrative calculations involve a sample mode
with $\ell=951$, $n=2$ and $\nu _0=4.5243\;$mHz. The curves in
Fig.~\ref{figsimple} correspond to a simple velocity profile. In
particular, it was assumed that the profiles of both the
rotational and the meridional flows are linear in the entire
(narrow) cavity in which the mode is confined. As one can see from
Eq.~(\ref{dispr}) the non-modal change in wavelength directly
manifests itself in a variation of the effective frequency in
time. However, the problem we are considering here has one
specific feature: the modes are trapped in the acoustic cavity
between the upper and inner turning points. As a result, if the
modal frequency increases (decreases), when it propagates from the
upper turning point to the inner one, then the frequency decreases
(increases) when mode propagates in the backward direction.
Therefore, any observable changes in modal frequencies (on the
surface), can only arise when the net influence of the flow,
during the propagation of the mode in the inward and outward
directions, is nonzero. Hence, the nonzero residuals in
Fig.~\ref{figsimple} result from the asymmetry of the background
flow. The example shown in Fig.~\ref{figsimple} corresponds to a
case with a simple flow velocity profile: for this calculation we
assumed linear rotational and meridional flow profiles in the
entire cavity. This simple profile is appropriate for the modes
with very high angular degree. The total residual during the
observed modal lifetimes is thus the sum of the net frequency
changes appearing during each passage from the upper turning point
to the inner one and backward as shown in Fig.~\ref{figsimple}
(Panel~$A$). The curve $a$ (solid line) corresponds to the mode
with $k_{x}>0$ and $k_{y0}>0$ and the curve $b$ to the $k_x>0$ and
$k_{y0}<0$ (dashed line). The points $1a$ and $2a$ correspond to
the moments when the mode $a$ reaches its upper and inner turning
points, respectively. Similarly we have points $1b$ and $2b$ for
the curve $b$. The Panel~$B$ in Fig.~\ref{figsimple} illustrates
the changes in modal frequency for different ratios $k_x/k_y(0)$,
i.e.\ for different directions of the horizontal wavevector in the
initial momentum. The curves on this panel in fact are the lines
connecting the residuals in frequencies corresponding to the
moments when the considered mode is close to its upper turning
points. We denote the curves for modes with $k_{y0}>0$ by the
characters with the index $1$ and for the modes with $k_{y0}<0$ by
characters with index $2$. If we represent the $k_{y0}$ components
of the wavevector as $k_{y0}=k_{h}\sin(\beta)$, where $\beta$ is
the angle between the $\vec k_h$ and $x$-axis, then the curves on
Panel~$B$ correspond to the values of $\beta$ in the following
order: curves $a_1$ and $a_2$ (crossed lines) correspond to
$\beta=\pm 5^{\circ}$, $b_1$ and $b_2$ (dashed lines) correspond
to $\beta=\pm 30^{\circ}$, $c_1$ and $c_2$ (solid lines)
correspond to $\beta=\pm 45^{\circ}$, $d_1$ and $d_2$ (doted
lines) correspond to $\beta=\pm 60^{\circ}$ and $e_1$ and $e_2$
(dash-dotted lines) correspond to $\beta=\pm 85^{\circ}$. Clearly,
the residuals in modal frequency due to the non-modal effects are
greater for modes propagating in an oblique direction with respect
to the both rotational and meridional flows and the effect is
maximal for the angle $\beta =\pm 45^{\circ}$ (i.e.\ $k_x=k_{y0}$)
(the angle corresponding for maximum rates of the frequency
residuals can differ from this, say it can be $\pm30^{\circ}$ or
$\pm60^{\circ}$, depending of values of shear rates and wavevector
components).

 In Fig.~\ref{figsimsmall} we show results
for a rather small value of the parameter $a$ ($a=-0.1\cdot
10^{-6}$ s$^{-1}$) obtained for the same mode as in
Fig.~\ref{figsimple}. The curves in panel $A$ correspond to the
angle $\beta=45^{\circ}$. The solid curve corresponds to the shear
rate $c=2.0 \cdot 10^{-5}\,$s$^{-1}$ and the dashed line
corresponds to $c=3.0 \cdot 10^{-5}\,$s$^{-1}$. Similarly, we show
the results for the angle $\beta=30^{\circ}$ in panel B, where the
dashed-dotted curve corresponds to the smaller value of parameter
$c$ and the dotted line to the larger one. It should be noticed
here that from both panels $A$ and $B$ one can easily see that the
amplitude of the frequency variation along the ray path increases
as $c$ increases. Finally, we combine the results for both angle
values $\beta=45^{\circ}$ (solid and dashed lines) and
$\beta=30^{\circ}$ (dashed dotted and dotted lines) calculated for
a 10 hour period. In this panel the line styles correspond to the
same values of the shear parameter $c$ as in the previous panels.
From this panel we conclude that the rate of the overall frequency
residuals drops down by a factor of 10 compared with the case
shown in Fig.~\ref{figsimple} since now the value of parameter $a$
is 10 times smaller. This seems to indicate that the frequency
residual rates depend nearly linearly on the value of this
parameter. We will come back on this point in the next section.

The aim goal of this subsection was to outline the principle of
the process and to get a rough estimate of the possible frequency
residuals for different  parameter values. For these purposes we
adopted a simplified velocity profile with fixed constant values
of the shear parameters within the entire cavity (above the
fractional radius 0.997), in which the examined oscillation mode
with $\ell = 951$ is confined.  This simple model yields the
following main conclusions: (1)~The joint effect of the
non-uniform rotational and meridional flows can yield finite
differences between the normal mode frequencies and the `effective
frequencies'; (2) These `residuals' can be of the order of a few
tens of $\mu$Hz. It is thus worthwhile to improve the model in an
attempt to better `quantify' the effect of the shear flows. This
is done in the next section, where we consider a more advanced
model including observed (not simplified) velocity profiles, and
perform calculations for different sample modes including the one
considered in the current subsection.

\section{Numerical results for observed velocity profiles}
In this Section we consider cases of modes with different angular
degrees. For these purposes, in general, more complicated, the
observed velocity profiles should be taken into account in the
model. Modes with $\ell$ significantly lower than that considered
in the previous subsection, penetrate deeper in the solar interior
and correspondingly we expand the width of cavity, where we
consider the propagation on modes as it is shown in Figures
\ref{prof} and \ref{proftemp2}. Further we explain the model in
detail.
\subsection{Modelling the subsurface flows}
As a basis for the present study we have the results of
helioseismological measurements of the subsurface velocity
profiles (see e.g.\ Gonz\'alez Hern\'andez et al.\
(\cite{gprdia}), Gonz\'alez Hern\'andez \& Patr\'on
(\cite{gons})).  These observations show that the gradients of the
flow velocity components are functions of position. That is the
reason why we present our results for a number of magnitudes of
the shear parameters confined within some ranges of values. The
model `shear parameters' included in the shear matrix are
estimated as follows. The photospheric latitudinal variation of
the rotation rate is usually modelled as:
\begin {equation}\label{rotlat}
\Omega (\theta) = \sum _{k=0,2,4}a_k\cos^k (\theta)+s_0,
\end{equation}
where $\theta$ denotes the colatitude and $a_0=452.0\;$nHz,
$a_2=-49.0\;$nHz, $a_4=-84.0\;$nHz, and $s_0=-31.7\;$nHz
(Snodgrass (\cite{sno}), Gonz\'alez Hern\'andez \& Patr\'on
(\cite{gons})). As the parameter $a=V_{0x,y}$ included in the
shear matrixes corresponds to the derivative with respect to the
$y-$coordinate of the $x$-component of the equilibrium flow
velocity. Thus, this parameter represents the latitudinal
variation of the rotational velocity in the considered local
plane-parallel slab. We performed the calculations assuming that
this parameter is of the order of $a \simeq -0.01\cdot 10^{-6}$, $
-0.1\cdot 10^{-6}$, $-0.3\cdot 10^{-6}$, $-0.6\cdot 10^{-6}$, or
$-1.0\cdot 10^{-6}$~s$^{-1}$.

\begin{figure}
\includegraphics[width=8cm]{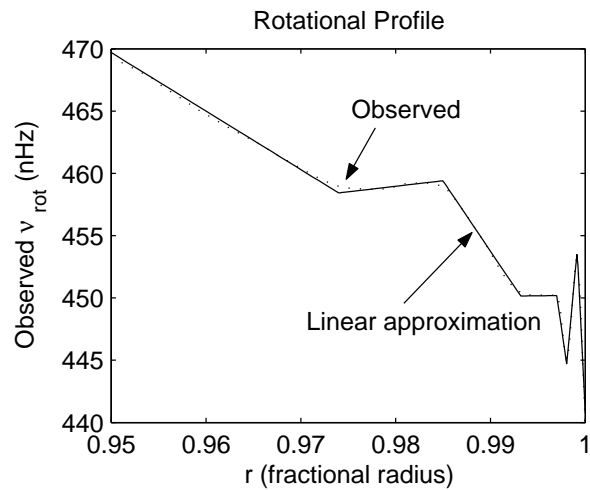}\\
\caption{Profile of the rotational velocity in the solar
subsurface layers. The solid line shows the sample profile for
latitude angle $15^{\circ}$ given by Gonz\'alez Hern\'andez \&
Patr\'on (\cite{gons}). The dotted curve is the observed
rotational profile. The solid lines show the modelled linear
approximations of the observed profile for different fractional
radii.}\label{prof}
\end{figure}

\begin{figure}
  \includegraphics[scale=1]{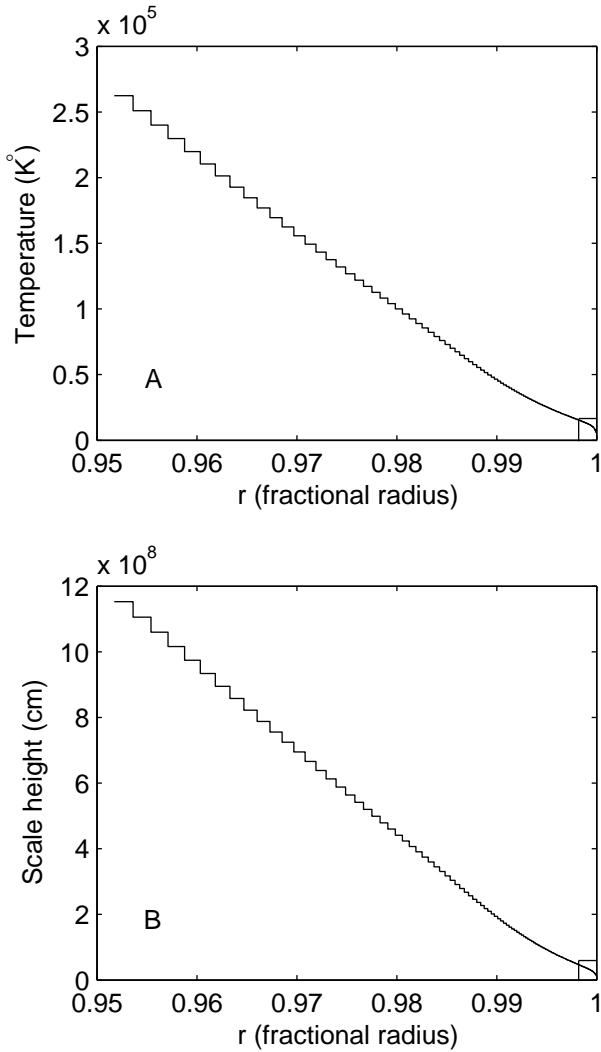}\\
  \caption{As in Figure \ref{proftemp1} corresponding to
the range of the fractional radius values $r=0.95 -1.00$. The
squares in the right bottom corners show the areas plotted in
respective panels of the Figure \ref{proftemp1}.}\label{proftemp2}
\end{figure}

To estimate the values of the parameter $b=V_{0x,z}$, which
locally represents the radial variation of the rotational
component of the flow velocity, and the radial shear $c$ of the
meridional ($y-$) component of the velocity field, we used results
of helioseismic inversions given in (Gons\'alez Hern\'andez et
al.\ (\cite{gprdia}), Gonz\'alez Hern\'andez \& Patr\'on
(\cite{gons})). In Fig.~\ref{prof} we show the rotational profile
for latitude $15^{\circ}$ taken from Gonz\'alez Hern\'andez \&
Patr\'on (\cite{gons}) (dotted line). We have made a linear
approximation of the observed profile. The approximate profile is
also plotted in Fig.~\ref{prof} (solid line). This approximate
profile corresponds to specific values of the shear parameter $b$.
The results of our estimates, for different fractional distances
from the solar center, are shown in the Table~1. At depths below
the thin sub-photospheric layer, i.e.\ for $r<0.95 R_{\sun}$, we
assume the radial gradient of the rotational velocity to be
negligibly small. Here it should be noticed that we only consider
oscillation modes with angular degree $\ell \geq 95$, which have
their lower turning points far above the solar tachocline. These
parameters $a$, $b$, and $c$, determine the inhomogeneity of the
background flow. Together with the spatial scales of the waves
themselves and with the timescales of the interaction between the
modes and the flows, they determine the strength of the influence
of the inhomogeneous flow on the waves.
\begin{table}
\centering \caption[]{Values of the shear parameters $b$ and $c$
at different fractional radii} \vspace{0.1 cm}
\begin{tabular}{cr|cr}
  \hline
  \hline
  $r/R{\sun}$  & $b$ & $r/R{\sun}$ & $c$ \\
   & $\times 10^{-6}$ s$^{-1}$ &  & $\times 10^{-6}$ s$^{-1}$\\
  \hline
  $\leq 0.9500$ & $0.0000$ &$\leq 0.950$& $0.0000$\\
  $0.9500-0.9740$ & $-2.5558$ &$0.95-0.971$& $ 2.2568$\\
  $0.9740-0.9850$ & $4.7295$ &$0.971-0.979$& $3.1034$\\
  $0.9850-0.9932$ & $-6.0004$ &$0.979-0.983$& $1.1111$\\
  $0.9932-0.9970$ & $0.0453$ &$0.983-0.988$& $8.0645$\\
  $0.9970-0.9981$ & $-27.420$ &$0.988-0.992$& $-12.9633$\\
  $0.9981-0.9992$ & $42.5677$ &$0.992-0.996$& $-16.6667$\\
  $0.9992-1.0000$ & $-80.4253$ &$0.996-0.998$& $-14.2857$\\
  $    -     $ & $      $ & $0.998-0.999$ & $21.2500$\\
  $    -     $ & $       $ & $0.999-1.000$& $28.5714$\\
  \hline
\end{tabular}
\end{table}

\subsection{Determination of the timescales of influence}
Another important issue we have to address concerns the timescales
of the interaction between the flow and the oscillation modes. One
of the basic temporal parameters characterizing the solar
$p$-modes concerns their lifetimes. Several observational methods
have been used by numerous authors to estimate the lifetimes
$T_{L}$ of the solar $p$-modes, see e.g.\ the work of of Chou et
al.\ (\cite{cho}) and Chen et al.\ (\cite{che}). There exists a
discrepancy between the observational measurements of the modal
lifetimes. Nevertheless, from the above-mentioned measurements one
can estimate the range of the lifetimes as $2$--$10$ hours.

\begin{figure*}
\centering
  \includegraphics[scale=0.77]{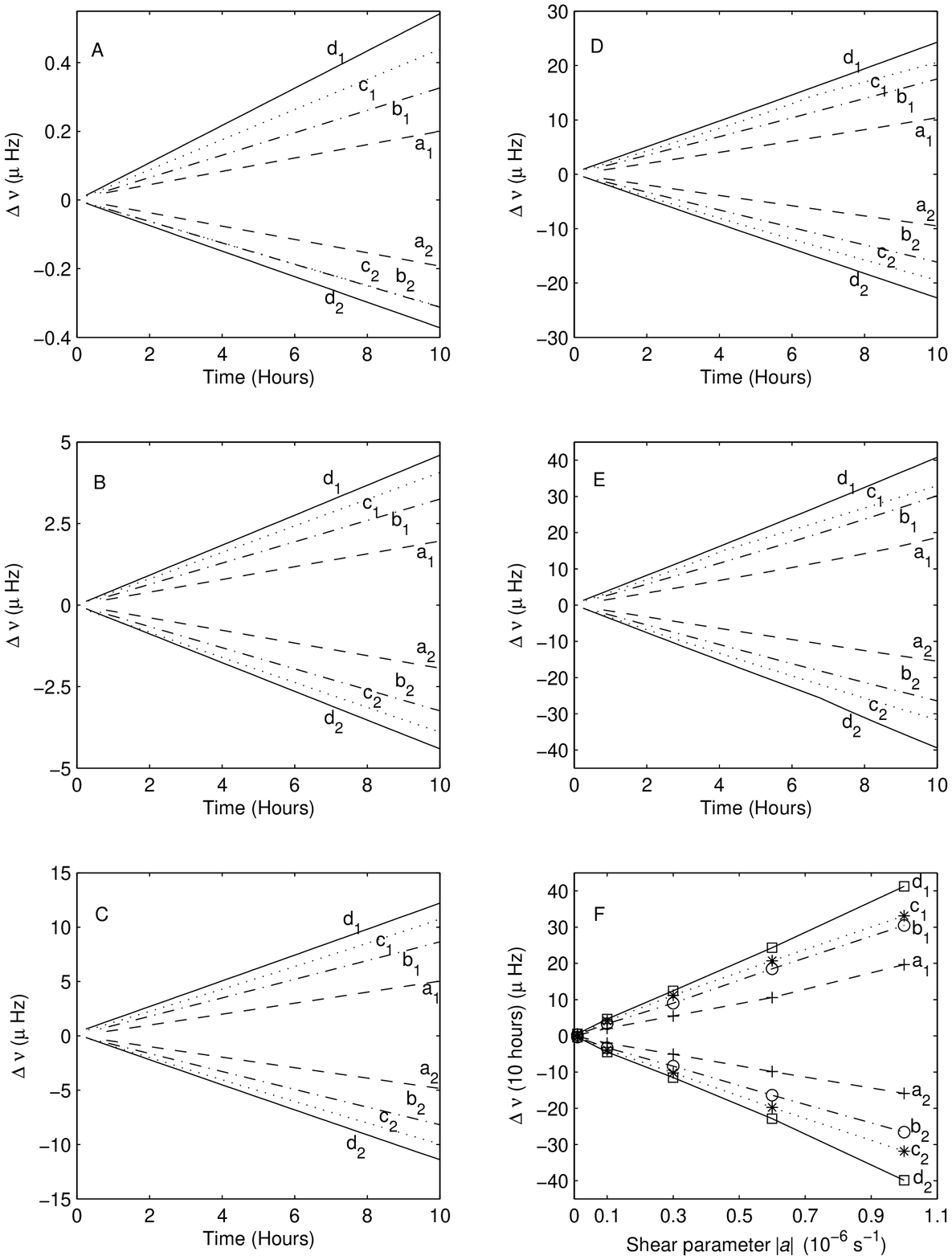}
  \caption{The variation of the modal frequencies in time calculated in the
case of the velocity shear profile described in Section~4.1 (see
Table~1). The direction of the horizontal wavevector is taken as
$\beta= \pm 45^{\circ}$. Curves $a_1$ and $a_2$ (dashed lines)
correspond to the sample mode with $\ell=95$, $n=4$ and with basic
modal frequency $\nu _0=2.3761$ mHz. Similarly, the $b_1$-$b_2$
(dashed dotted lines) and $c_1$-$c_2$ (dotted lines) respectively
are curves for the modes $\ell=263$, $n=6$, $\nu _0=4.2336$ mHz
and $\ell=673$, $n=2$, $\nu _0=3.8202$ mHz. And the curves
$d_1$-$d_2$ (solid lines) correspond to the same mode with $\ell
=951$ considered in section 3.2. Different panels show results of
calculations in the three different cases of the latitudinal
gradient of the rotational velocity: Panel $A$: the parameter
$a=-1\cdot 10^{-8}$ s$^{-1}$; panel $B$: $a=-1\cdot 10^{-7}$
s$^{-1}$; Panel $C$: $a=-3\cdot 10^{-7}$ s$^{-1}$; Panel $D$:
$a=-6\cdot 10^{-7}$ s$^{-1}$; Panel $E$: $a=-1\cdot 10^{-6}$
s$^{-1}$. Note that the ordinate axes in panels A-E have different
scales. In panel~F the frequency residuals arising after a 10 hour
period are plotted as functions of the shear parameter $|a|$
absolute values.}\label{figgen}
\end{figure*}

Another characteristic temporal parameter of an oscillation mode
is its travel time $\tau$ along the ray path, from one reflection
in the upper turning point to another one (see Fig.~1).  As it is
well known these travel times are observable by applying the
methods of local helioseismology.

It is well-known that the radial gradient of the temperature and,
therefore, of the sound speed inside the Sun leads to the
refraction of waves. Hence, in order to properly calculate the
changes in the modal frequencies as the modes propagate along
their ray path, we radially discretize the plain-parallel slab. We
thus consider a set of very thin shells so that within each of
these thin plasma layers the local sound speed, the cut-off
frequency, and the Br\"{u}nt-V\"{a}is\"{a}la frequency can be
assumed to be constant. Therefore, the spatial orientation of the
wavevector also approximately remains constant when the waves
propagate within the given layer. In our calculations we use the
equilibrium parameters taken from the standard solar model (see
e.g.\ {\tt www.ap.stmarys.ca/$\sim$guenther/solar/ssm455.sink}).
We denote the part of the total ray path confined within the given
layer by $ds={v_{ph}} dt$ where, $v_{ph}=\omega /k$ is the modal
phase velocity and $dt$ is the time interval during which the wave
front stays within the given layer. This distance relates with the
distance between the top and bottom boundaries of the layer $dz$
as follows:
\begin{equation}\label{dz}
    dz=ds \cos \alpha,
\end{equation}
where $\alpha$ is the angle between the vector ${\vec ds}$ and the
vertical (see the schematic view in Fig.~\ref{f1}). The cosine
of this angle can be written in terms of the components of the
wavevector. In particular, $\cos \alpha =k_z/k$ where $k_z$ and
$k$ are the radial component and length of the wavevector,
respectively. Finally according to the above expressions we can write:
\begin{equation}\label{dt}
    dt = \frac {k^2}{\omega k_z} dz.
\end{equation}
This last expression determines the time interval during which the
wave undergoes an influence from the inhomogeneous flow within
the given thin layer. Hence, in our model we consider the
influence of the inhomogeneous flows on the oscillation modes as
{\em the accumulation of the effects in the different layers reached by
the given mode as it propagates from its upper to its lower
turning point and back}. Now, as we have parameterized the
effective time of influence of the nonuniform flows on the
$p$-modes, the question that remains is: ``how strong is this
influence?''. In other words, is the total time interval of influence
large enough for significant changes in the modal properties or not?
Below, we concentrate on this issue.

\subsection{Modes with angular degree $\ell < 200$}
In this subsection, results  of calculations are presented based
on the model discussed in the previous subsections, i.e.\ the more
general case with realistic approximations of the velocity
profiles. The shear parameters used for these calculations depend
on depth and are given in Table~1. In Fig.~\ref{figgen} the
results of these calculations are shown for different values of
the shear parameter $a$. The panels $A$, $B$, $C$, $D$ and $E$
correspond to $\beta=\pm 45^{\circ}$ and to the values $a=-1\cdot
10^{-8}\;$s${-1}$, $a=-1\cdot 10^{-7}\;$s${-1}$, $a=-3\cdot
10^{-7}\;$s${-1}$, $a=-6\cdot 10^{-7}\;$s${-1}$ and $a=-1\cdot
10^{-6}\;$s${-1}$, respectively. Again the curves labelled by
characters with index $1$ correspond to modes with $k_{y0}>0$,
while the labels with index $2$ correspond to modes with
$k_{y0}<0$. The curves $a_1$ and $a_2$ (dashed lines) correspond
to the mode with angular degree $\ell=95$, $n=4$ and a basic modal
frequency $\nu _0=2.3761\;$mHz. Clearly, the influence of shear
flow is more pronounced as time proceeds. It turns out to yield
changes in the frequency of the order of a few $\mu$Hz to, at
most, 5 to $20\;\mu$Hz after 10 hours (the maximal life time of
the mode). The values of the shear parameter $a$ used in the
calculations correspond approximately to the latitudes
$1^{\circ}$, $10^{\circ}$, $25^{\circ}$, $40^{\circ}$ and
$60^{\circ}$, respectively. The overall frequency residual (after
10 hours) for the considered mode is plotted in panel~F of
Fig.~\ref{figgen} (dashed line) as a function of the parameter
$a$.

\subsection{Modes with angular degree $\ell \geqslant 200$}
From the Eqs.~(\ref{eqksol}) and (\ref{dispr})-(\ref{W}) it is
clear that the rate of change of the modal wavelength in time
strongly depends on the modulus (and consequently on the angular
degree of modes as $k_h ^2 \propto \ell (\ell +1)$) and the
direction of the wavevector, i.e.\ on its horizontal ($k_h$) and
radial ($k_z$) components. In particular, the larger the component
$k_{h}$ of the wavevector is, the more substantial the changes of
the modal properties are. Therefore, we can expect that the
deviation of the modal properties (from those of normal modes)
increase with angular degree. As a matter of fact, our
calculations show larger residuals from the basic modal frequency
for the sample modes (see Fig. \ref{figgen}): $\ell=263$, $n=6$,
$\nu _0=4.2336\,$mHz (curves $b_1$ and $b_2$ (dash-dotted lines)),
$\ell=673$, $n=2$, $\nu_0=3.8202\,$mHz (curves $c_1$ and $c_2$
(dotted lines)), and the mode considered in section 3.2 (curves
$d_1$ and $d_2$ (solid lines)). These plots demonstrate the fact
that the residuals of the modal frequency due to the non-modal
effects increase with growth of the angular degree. The overall
residuals after 10 hours versus parameter $a$ are also plotted for
these sample modes in panel~F in the respective order of line
styles. This latter confirms that the frequency residuals depend
on the values of the parameter $a$, and this dependence has a
nearly linear character.

\begin{figure*}
  \centering
  \includegraphics[width=12cm]{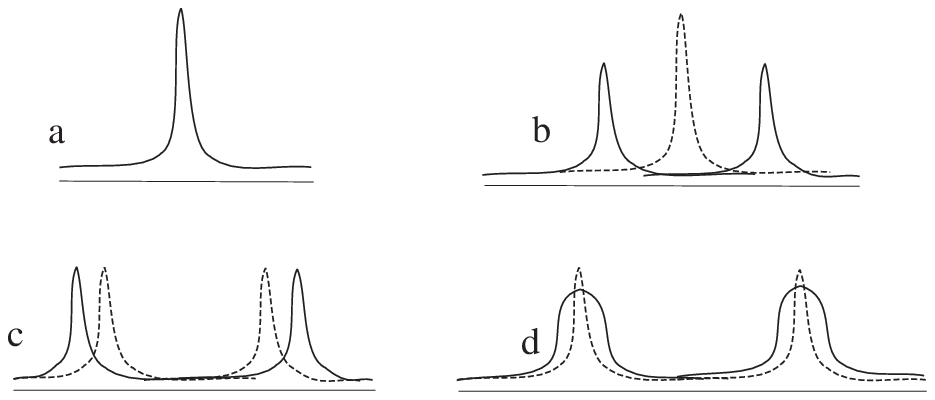}
  \caption{The schematic view of the peak(s) in the power spectrum corresponding
to the oscillation mode with a given angular degree. a)~In the
static medium b)~in the case of uniform flow; c)~Case of the
inhomogeneous flow without the non-modal effects; d)~Case of the
inhomogeneous flow including the non-modal
effects.}\label{figspect}
\end{figure*}

In addition, it should be noticed that the modes undergo a more
substantial influence of the flow when they are closer to their
inner turning points. As a result, modes with inner turning points
within the area with strongly pronounced gradients of the flow
velocity (i.e.\ mostly the modes with $\ell\geqslant 200$), change
in time more effectively. This fact also contributes to the
appearance of more significant non-modal effects for modes with
higher angular degree. Comparing the results shown in
Fig.~\ref{figgen} for the sample mode with $\ell=951$ (solid lines
in panels~B and E) with those obtained with the simplified
velocity shear profiles (see panel B in Fig.~\ref{figsimple} and
panel C in Fig.~\ref{figsimsmall}) it is easily seen that our
preliminary estimates given in section~3.2, yielding that
frequency residuals up to values of a few tens of $\mu$Hz should
be expected, were true. For $a=-1\cdot 10^{-7}\;$s${-1}$ the solid
line d1 in panel~B in Fig.~\ref{figgen} should be compared with
the solid and dashed lines in panel~C of Fig.~\ref{figsimsmall},
which correspond to the angle $\beta=45^{\circ}$. On the other
hand, from the panel~E of Fig.~\ref{figgen} one can see that
(comparing the solid curves d1 and d2 with the similar curves in
panel~B of Fig.~\ref{figsimple}) the characteristic frequency
residuals with the observed velocity profiles are somewhat smaller
than those obtained with the constant shear rates in subsection
3.2. This is due to the complicated radial profile of the observed
rotational and meridional flows leading to alternating signs of
the shear rates along the radial direction.

\section{Discussion}
Formally speaking, non-uniform flows can influence the linear wave
modes in fluids and plasmas in two different ways. On one hand the
flow advects the wave front producing a Doppler shift of the modal
frequencies (the scalar product on the RHS of the
Eq.~(\ref{dispr})). On the other hand, the shear background flow
causes (in general) changes in the modal wavelength, which in turn
results in a temporal variation of the characteristic frequencies
and/or amplitudes of the linear disturbances (for the case of
modes considered here see the last term in RHS of
Eq.~(\ref{dispr})). Under the standard normal mode formalism it is
{\it a priori} assumed that the perturbations evolve as purely
exponential ($\sim \exp (i\omega t)$) functions of time. As it is
well known, the Doppler shift of the modal frequencies can be
described under the normal mode analysis very well. On the other
hand, as we have shown the temporal changes in wave
characteristics related to the time dependence of the wavenumbers
can be studied as well. The normal mode decomposition of the
perturbations neglects this latter characteristic of the temporal
evolution of the wave modes. For the particular case of $p$-modes
trapped in the solar interior this assumption is true in most
cases. Indeed, there are no sharp flow velocity gradients observed
in the solar interior. Even in narrow shells below the convection
zone and just beneath the photosphere, where significant gradients
of the flow velocity are observed, the shear rates still are by
several orders of magnitude lower than the characteristic
frequencies of the oscillation modes. Therefore, we expected that
the inhomogeneous flows in the subsurface shell, which is under
consideration here, could only slightly deviate (non-modally) the
temporal behavior of the modes from that described under the
standard normal mode approach. The approximative analysis
performed in this work leads to the conclusion that these
expectations were true. Nevertheless, the considered non-modal
effects still turn out to be {\lq}strong{\rq} enough to support
different observational evidences. In particular, from the global
point of view the non-modal effects can cause the {\it `blurring'}
of the global $p$-mode power spectra and the mixing of the modal
power with the noise. On the other hand, these effects contribute
to a slight {\it `deformation'} of the modal ray paths. We address
these observational aspects of the non-modal effects in details in
following subsections.

\begin{figure*}
  \centering
  \includegraphics[width=14cm]{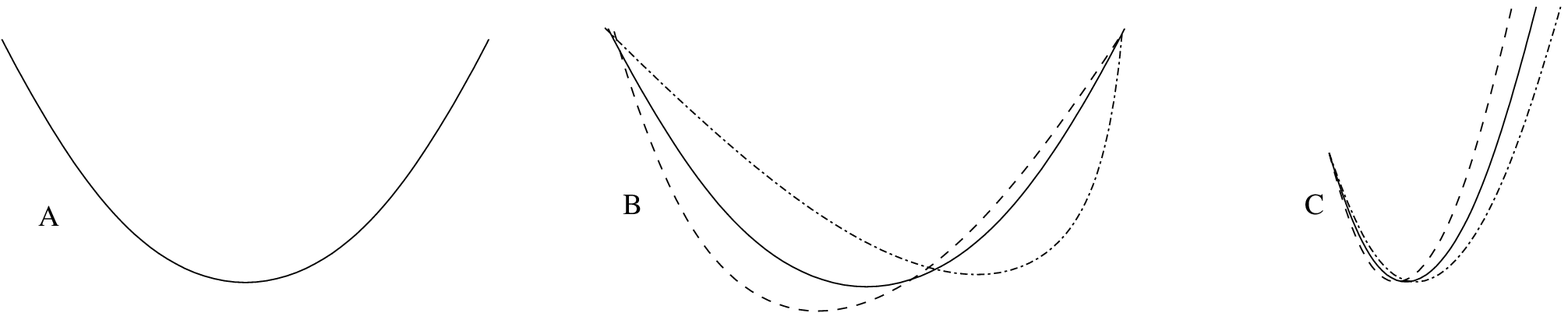}
  \caption{The schematic view of the modal ray paths: A) without non-modal effects;
B)Same as on panel A (solid line) and the non-modal deformation of
the ray path due to the temporal evolution of the radial component
$k_z$ of the wavevector, which first increases and then decreases
in length (dashed line) and vice versa (dash-dotted lines); C) The
non-modal deformation of the ray path because of the temporal
evolution of the meridional component $k_y$ of the wavevector
(dashed and dash-dotted lines). The modes do not stay in one
plane, but the trajectories become 3D curves.}\label{figlocal}
\end{figure*}

\subsection{The observational aspects of the non-modal
effects - global view}

Let us now address the important issue about how the effects
related to the flow velocity gradients manifest themselves in the
observed power spectrum of the solar $p$-modes. It is believed
that the solar oscillations are excited by the stochastic motions
in the upper part of the convection zone. From the global point of
view, the solar $p$-modes are observed as global oscillation modes
of the Sun by measuring the characteristic frequencies of the
intensity or the velocity field oscillations. The observational
data are then concentrated in the power spectrum which enables the
identification of different oscillation modes by fitting the peaks
arising in the spectrum with the peaks in the analytically
obtained Fourier power spectrum. The formation of each peak
involves the contribution of many modes with similar properties.
The oscillation modes are characterized by their frequency, their
radial order $n$, their angular degree $\ell$, and their azimuthal
order $m$. As it is well known from the theory of stellar
non-radial oscillations (Unno et al.~\cite{unno}), in the case of
a spherically symmetric star the eigenfrequencies are degenerate
with respect to the azimuthal order. This means that it is not
possible to distinguish (in the observed power spectrum) modes
with the same radial order and angular degree but propagating in
different horizontal directions. In Fig.~\ref{figspect}, panel (a)
schematically shows one peak corresponding to a given radial order
and angular degree and containing the power of all modes with
azimuthal order $m=-\ell,-\ell+1, ..., 0, ..., \ell-1,\ell$.

Any phenomenon that spoils the spherical symmetry (for example the
axi-symmetric rotation of the star, magnetic field etc.) results
in lifting the degeneracy with respect to $m$. In the
non-spherically symmetric case, instead of one peak their appear
several peaks in the power spectrum corresponding to the different
orientations of the horizontal wave vector. On panel (b) of
Fig.~\ref{figspect} this is illustrated schematically by two
peaks in the power spectrum corresponding to two modes with the
same radial order and angular degree, but with horizontal wave
vectors directed to opposite directions. This effect is the
well-known phenomenon of `frequency splitting'. Clearly, frequency
splitting observations allow to perform the inversion of the
observational data into the rotational profile inside the Sun. It
should be noticed, however, that the eigenfrequency splitting can
also be caused by any other factor breaking the spherical symmetry
such as, for example, the meridional flows.

The effect of inhomogeneous flows on the power spectrum can be
complicated. On one hand, the flow velocity gradients contribute
to the frequency splitting just like the uniform flow would do. On
the other hand, we have shown above that the influence of the
flow inhomogeneity can be noticeably important, able to serve
as a mechanism causing  changes in the modal properties which
can contribute to the observed deficit of modal power for very
high degree $p$-modes. In the non-modal approach we have described
these changes by a time dependent wavevector and a time dependent
modal frequency. Hence, to describe the effect of inhomogeneous
flows on the power spectrum we need to use additional
characteristic temporal parameters: viz.\ the widths of the peaks
in the power spectra $\Delta \nu$ and the observation time, which
relates to the resolution of the observations. If the changes in
the modal frequencies $\delta \nu$ occurring due to the non-modal
effects during the lifetime of the mode are much smaller than the
half-width of the corresponding peak in the power spectrum, then
one can assume that the changes in frequency are negligible. In
this case the background flow only contributes to the observed
frequency splitting. This is the case considered under the
formalism of the standard modal analysis by Ulrich et
al.~(\cite{ulr}). The schematic view of this situation is given on
panel (c) in Figure~\ref{figspect}.

When $\delta \nu \simeq \Delta \nu$ the changes in frequency
contribute to the line width (along with different damping
mechanisms causing a variation of the mode amplitudes). When
$\delta \nu > \Delta \nu$ the frequencies of the individual modes
are 'drifting' along the frequency axis. In this case a part of
their power is mixed with the acoustic noise (leading to a
decrease of the signal to noise ratio) and a part of their power
contributes to the neighboring peaks. Under these circumstances,
even in the conservative case where we assume that the average
power exchanged by two neighboring peaks are approximately equal
to each other (in the real situation this can not be true, see
Goldreich et al.~(\cite{gomur})), the non-modal effects could be
the cause of the partial dissipation of the power because of the
mixing with noise. The last two situations are schematically
displayed on panel (d) in Fig.~\ref{figspect}. Here we should make
one important remark. The modes that contribute to a given peak in
the power spectrum are randomly excited by different convective
sources at different time intervals. But, as the flow velocity
gradients are small on the Sun the changes in frequencies of the
particular mode by the non-modal effects need a significant time
to occur. That is why the peaks in the power spectrum remain to be
concentrated around the basic central frequency and why the
frequency splitting is still observable.

\subsection{The observational aspects of the non-modal
effects - local view}
Now we turn to another issue, viz., how the non-modal changes in
modal properties affect the observational data obtained by using
the well-known methods of local helioseismology (such as
time-distance techniques, ring diagrams, etc.). In fact, the
expressions (\ref{eqph}) determine the shape of the trajectory of
a point with a constant phase, within the framework of our model
with a piecewise constant temperature. One can easily see from
these expressions that in the shearless (zero shear matrix) limit
the horizontal components ($k_x$ and $k_y$) remain constant.
Therefore, in this case the modes propagate in the horizontal
direction along straight lines ($x$ and $y$ coordinates of the
point are linear functions of time) and the ray path is confined
in one plane (the problem is two dimensional). The variation of
the radial component of the wavenumber $k_z$ arises from the
refraction of waves by the temperature gradient along the radius.
The schematic view of a given modal ray path between two
neighboring reflections at the upper turning points is shown in
Figure~\ref{figlocal} (panel A). The effects which can
systematically distort the ray paths (when we exclude non-modal
changes) relate only to the inhomogeneous temperature and/or
magnetic field profiles and Doppler shift of frequencies due to
the advection of modes by the flows (this effect is represented by
the scalar product on the RHS of Eq.~(\ref{dispr})). These
systematic distortions are immediately detectable by using local
methods of helioseismology. On the one hand, observations detect
changes in modal travel times and distances. On the other hand,
changes of modal frequencies result in a transformation of the
so-called acoustic ring shape from a ring to an elliptical one.

If we turn now to the case of the nonzero shear rates, it can be
seen from Eq.~(\ref{eqph}), that the components of the wavevector
are time dependent, in general. The character of this dependence
is determined by the profiles of background non-uniform velocity
field. In the particular case, which we consider in this paper
(rotational velocity depends on $y$ and $z$ coordinates and the
meridional flow varies only radially) we get the `dispersion
relation' (\ref{dispr}) through introducing an `effective
frequency' because of the rather small shear rates and,
consequently, a very slow and small variation of the wavenumbers
and frequencies in time. The non-modal changes of the `effective
frequency' and wavevector result in a slight deformation of the
ray path. In particular, the radial component of the wavevector
$k_z$ now varies along the ray path because of two reasons. The
first one is the usual variation due to the temperature gradient.
The second reason is related to the non-modal variation (see
Eq.~(\ref{skzsf})). If $k_z$ increases (decreases) within a given
layer, when the mode propagates from the upper turning point to
the lower one, then the non-modal variation of the radial
wavenumber turns and extends (shortens) the wavevector. The
wavevector behaves oppositely when the mode propagates from the
lower turning point to the upper one: within the same layer it
turns in the opposite direction and its length respectively
decreases (increases) resulting in a compensation of the residual
of the `effective frequency' produced within the same layer
before. Therefore, the combined action of the refraction and
non-modal effects produces the shape of the ray path in the
$k_{h0}-k_{z0}$ plane schematically shown in Figure~\ref{figlocal}
(panel B).

It should be noticed that the uncompensated finite residuals in
modal wavelength (frequency) can arise only in the case of nonzero
$a$ and $c$ coefficients. This is the reason why one can claim
that the slight non-modal evolution of the $p$-modes occurs only
under the joint effect of the gradients of the rotational
(parameter $a$) and meridional (parameter $c$) flows. The nonzero
value $a$ leads to the variation of the $k_y$ (Eq.~(\ref{skysf}))
in time. Because of this the ray paths do not stay in the
$k_{h0}-k_{z0}$ plane, but instead they become 3D curves. The
schematic view of these curves are shown in Figure~\ref{figlocal}
(panel C).

The residuals occurring during two neighboring reflections are
very small and they can not be detected in the observational data
of the modal travel times and distances. The non-modal variation
of the modal properties is a cumulative effect in the case of the
trapped modes occurring only after a significant number of
reflections, while the effect related to the Doppler shift of the
frequencies is immediately detectable as it systematically shifts
all frequencies at once. Formally speaking the non-modal variation
of the frequency is a `second order' effect and it only becomes
significant after some time after the modal excitation. The effect
considered in this paper can not be detected by ring diagram
methods either. As it is well known, the acoustic rings represent
slices of the 3D local power spectrum at a given fixed frequency
and deformation of the ring demonstrates the shift of frequencies
due to advection (for example see Hill (\cite{hill})). While
because of the non-modal variation the `effective frequency'
slowly migrates along and/or across the corresponding ridge. That
is why we do not expect that the non-modal variation will cause
any additional deformation of the slices of the spectrum by fixed
frequency planes. However, some signature of the non-modal
variation can arise because of the azimuthal variation of the
height (or width of the ridges). But this issue could become the
subject of a separate study in order to verify whether the order
of magnitude of the mentioned azimuthal variation is compatible
with the resolution of the employed method (in this case a FFT).
To detect the non-modal effects efficiently it is convenient to
develop a different kind of multidimensional spectral method
involving different methods of data analysis with variable
temporal and spatial scales (such as a window fourier transform or
a wavelet analysis). This latter issue is out of the scope of
current work and could be a matter of future studies.
\section{Conclusions}
The influence of the observed subsurface flow inhomogeneity on
solar $p$-modes has been studied in the framework of a {\em
non-modal analysis}. Particular attention was given to the
possible role of the non-modal time-dependent effects in the
formation of the observed $p$-mode power spectra. We suggest the
hypothesis that several properties of the high-degree $p$-modes
could be attributed to the slight (`non-modal') deviation of the
temporal behavior of this kind of modes from a purely (`modal')
exponential ($\sim \exp (i\omega t)$) evolution. This effect is
related with the non self-adjointness of the governing equations
which leads to the existence of alternative solution of these
equations describing the temporal evolution (in general
non-exponential) of perturbations. This statement is based on the
fact that high degree $p$-modes are mostly confined in the upper
thin subsurface shell with strongly pronounced inhomogeneous
flows. We examined this hypothesis and, in particular, it was
found that:
\begin{description}
\item{a)} It is a well-established observational fact that the
line widths of the peaks in the $p$-modes power spectrum increase
with angular degree of modes. In the previous section we have
discussed the observational aspects of the flow inhomogeneity
related phenomena. Our results show that the nonuniform flows can
contribute to the excess line widths (compared to the inverse
observational time) in the observed power spectrum. Clearly, the
effect addressed in this paper acts along with other damping
mechanisms. As the frequencies in the spectrum are concentrated in
the peaks around some central frequencies the temporal rates of
stochastic excitement of the modes by different convective sources
are greater than the characteristic time of the significant
changes in the modal properties. That is why the effect discussed
here can not be observed as a shift of the peak in the power
spectrum. Although, some shift can appear because of the different
values of residuals in frequency of modes propagating in different
horizontal directions with respect to the meridional flow, but
that effect would be much less significant than the changes of the
particular wave frequency excited by a given convective source.
Another consequence supporting this scenario is that the residuals
appear due to the asymmetry of the considered velocity field. In
the case when the terms in the approximate `dispersion relation'
containing information about the meridional flow are negligibly
small, the modes undergo a much smaller net effect from the flow
non-uniformity during the propagation Sun in the inward and
outward directions.

\item{b)} We considered the effect of the inhomogeneous background
flows on the lower degree ($\ell< 200$) modes. These modes have
their turning points deeper in the convection zone, where we have
only latitudinal differential rotation. In this case, the effect
of inhomogeneous flows is of the order of 6-12 $\mu$Hz and this
fact is in agreement with (and it confirms) the observational
evidence that the theory of mode excitement/damping explains the
observed energetic spectrum very well for values of the angular
degree $\ell <200$.

\item{c)} Finally, the joint effect of the differential rotation
and the inhomogeneous meridional flow on the oscillation modes
with very high angular degree ($\ell \gg 200$) causes {\em
significant changes of the mode frequencies in time}. This may
mean that the evolution of the $p$-modes due to the background
velocity shear takes part in the distribution of the wave energy
over frequencies and angular degrees. Together with several other
possible damping mechanisms, this effect can explain or contribute
to the observed energetic spectrum of the modes that shows a
systematic deficit of the contributions from high degree modes.
This may be caused by partial mixing of power related to the given
mode with the acoustic noise. Hence, the suggested mechanism may
be the cause of the discrepancies between theoretical predictions
from the model of turbulent exitation/damping of oscillations on
the sun on the one hand, and helioseismic observations of very
high degree $p$-modes on the other hand. Additionally, the changes
of modal frequency due to the non-modal effects might be a
possible reason for the increases of the wide line widths of very
high degree solar $p$-modes.
\end{description}

The assumptions we made during the analysis of our theoretical
results enable us to study the main features of the modes
propagating in the nonuniform flow. But they only allow us to
understand  the `principle' of the process and to estimate the
possible changes of the modal properties related to the non-modal
effects. In order to study this problem quantitatively more
systematically the following efforts have to be made in the
future: (i) The analytical model should be further developed and
probably multidimensional spectra should be constructed (analogous
to the ring diagrams) involving also different methods of data
analysis (say wavelets); (ii) Direct numerical simulations of the
governing equations in real space and time should be carried out;
(iii) The results of these simulations should be analyzed and
compared with the new observational data.

\begin{acknowledgements}
The authors thank ~A.~Rogava for informative discussions and
suggestions. This work has been developed in the framework of the
pre-doctoral program of B.M.~Shergelashvili at the Centre for
Plasma Astrophysics, K.U.Leuven (scholarship OE/02/20). These
results were obtained in the framework of the projects OT/02/57
(K.U.Leuven) and 14815/00/NL/SFe(IC) (ESA Prodex 6). We express
our gratitude to Prof.\ K.G.\ Libbrecht for allowing us to use the
$p$-mode frequency data. We are also thankful to the anonymous
referee whose comments and suggestions led to the significant
improvement of the manuscript content.
\end{acknowledgements}

\end{document}